\documentclass[pre,twocolumn,twoside,superscriptaddress,amsmath,amssymb]{revtex4}

\usepackage{graphicx}

\newcommand{\abs}[1]{\left| #1 \right|}
\newcommand{\vect}[1]{{\bf #1}}
\newcommand{\uvect}[1]{{\bf \hat{#1}}}

\newcommand{\diss}{\lambda}
\newcommand{\tg}{\tau_c}
\newcommand{\dens}{\rho}

\newlength{\gxlen} 
\newlength{\hgxlen} 

\newcommand{\eref}[1]{Eq.~(\ref{eq:#1})}

\newcommand{\fref}[1]{Fig.~\ref{fig:#1}}
\newcommand{\Fref}[1]{Fig.~\ref{fig:#1}}
\newcommand{\ssref}[1]{\ref{sec:#1}}
\newcommand{\sref}[1]{section~\ref{sec:#1}}
\newcommand{\Sref}[1]{Section~\ref{sec:#1}}
\newcommand{\lqq}{``}
\newcommand{\rqq}{''}

\begin{document}

\title{Cluster Growth in two- and three-dimensional Granular Gases}          
\author{S.~Miller}
\affiliation{
  Institut f\"ur Computeranwendnungen~1, 
  Universit\"at~Stuttgart, Pfaffenwaldring~27, D-70569~Stuttgart, 
  Germany
  } 
\author{S.~Luding}
\affiliation{
  Institut f\"ur Computeranwendnungen~1, 
  Universit\"at~Stuttgart, Pfaffenwaldring~27, D-70569~Stuttgart, 
  Germany
  } 
\affiliation{
  Particle Technology, DelftChemTech, TU~Delft, 
  Julianalaan~136, 2628~BL~Delft, The~Netherlands\\
  e-mail: S.Luding@tnw.tudelft.nl
  } 
\date{\today}

\begin{abstract}
Dissipation in granular media leads to interesting phenomena
such as cluster formation and crystallization in non-equilibrium 
dynamical states.
The freely cooling system is examined concerning
the energy decay and the cluster evolution in time, both in two and three
dimensions.
We also suggest an interpretation of the 3D cluster growth 
in terms of percolation theory,
but this point deserves further study.
\end{abstract}


\maketitle

\newpage

\section{Introduction}

Granular media are interesting multi-particle
systems with a rich phenomenology 
\cite{herrmann98,cafiero00,poschel01,vermeer01,kishino01}. 
They can form a hybrid state between a fluid and a solid, where
the behavior is controlled by the balance between energy input
and energy dissipation. 
Energy input leads to a reduction of the density due to more collisions
and increasing pressure, so that the material can flow. 
In the absence of energy input, e.\,g., in 
freely cooling systems, granular materials 
become denser, i.\,e.\ they locally \lqq{}solidify\rqq{} 
due to dissipation.  
Because of mass conservation, the local densification is accompanied 
by a density decrease in other parts of the system, giving rise
to complex patterns and structures, with an interesting time evolution.
However, theoretical approaches are non-classical and appear often 
extremely difficult, so there is still 
active research directed towards the better understanding of granular 
media.

The subject of this paper is the pattern formation via clustering
in a dissipative, freely cooling granular gas 
\cite{goldhirsch93,goldhirsch93b,mcnamara96,luding99,nie02,das03a,das03b}.
The basic idea of clustering is that in an initially homogeneous freely 
cooling granular gas, fluctuations in density, velocity, and temperature 
cause a position dependent energy loss.
Due to strong locally inhomogeneous dissipation, pressure and energy drop 
rapidly and material moves from \lqq{}hot\rqq{} to \lqq{}cold\rqq{} regions, 
leading to even stronger dissipation and thus causing the density
instability with ever growing clusters 
until eventually, clusters reach system size (see \fref{snap}).

We investigate this clustering instability with respect 
to different dissipation rates and different system sizes.
Even though many of our findings are empirical, we attempt to reduce
the complexity of the evolution of the system by use of simple scaling
laws.

In \sref{details}, we explain in detail the simulation approach.
The results of our numerical experiments are discussed in \sref{numeric}.
Finally, in \sref{summary} a short summary and a discussion are given. 

\begin{figure}
  \begin{center}
    \includegraphics*[width=1.12\gxlen]{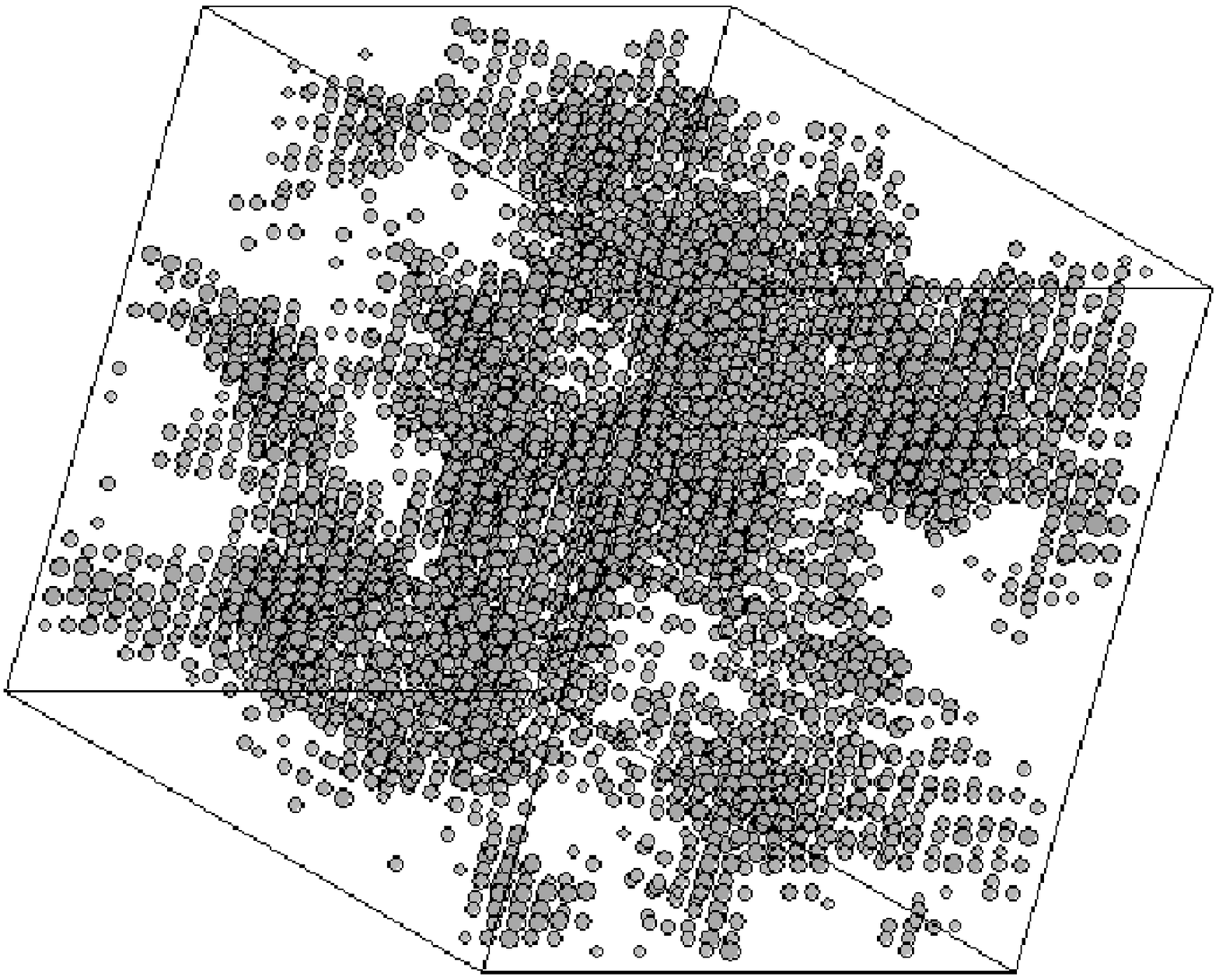}\\*
  \end{center}
    \caption{Density distribution in a snapshot of a 3D system 
      with 512000 particles, 
      volume fraction $\dens=0.25$, and restitution coefficient $r=0.3$.\\ 
      Each sphere represents a cell with about 40 particles in average.
      The size of the spheres is proportional to the local density; 
      very small spheres (corresponding to low densities) are omitted.
      }
    \label{fig:snap}
\end{figure}

\section{Simulation Details}
\label{sec:details}

A granular gas can be idealized as an ensemble of hard spheres 
in which the energy loss that accompanies 
the collision of macroscopic particles is modeled 
with a single coefficient of restitution $r$.
In the simplest case the particles are identical in size and mass 
and there are no inter-particle forces between collisions.

Details about initial and boundary conditions are given in
\sref{init}.
The microscopic dynamics of the motion and the collision of the particles is 
discussed in \sref{micro} and the simulation method is explained in \sref{ed}.
\Sref{tc} deals with the inelastic collapse, 
a problematic artefact of the hard sphere model with dissipation.

\subsection{Initial and Boundary Conditions}
\label{sec:init}

The simulation volume consists of a box with equal side length 
and periodic boundary conditions in two or three dimensions.

An initial state with random particle positions and velocities 
is prepared in the following way:
The particles first sit on a regular lattice and 
have a Maxwellian velocity distribution 
with a total momentum of zero. 
Then the simulation is started without dissipation 
and runs for about $10^2$ collisions per particle.
This state is now used as initial configuration for 
the dissipative simulations.

\subsection{Microscopic Dynamics}
\label{sec:micro}

Between collisions no forces act upon the particles 
and they move at constant velocity.

The particles are idealized as hard spheres.
This means that collisions take infinitesimal time
and involve only two particles.
Conservation of momentum leads to the collision rule
\cite{goldhirsch93,luding98cref}
\begin{align}
\vect{v'}_{1/2} &= \vect{v}_{1/2} \mp \frac{1+r}{2} 
\left( \uvect{k} \cdot ( \vect{v}_1- \vect{v}_2 ) \right) \uvect{k} \; ,
\end{align}
where a prime indicates the velocities $\vect{v}$ after the collision, 
$\uvect{k}$ is a unit vector pointing along the line of centers 
from particle 1 to particle 2, and $r$ is the coefficient of restitution.
The relative tangential velocity does not change during a collision,
the relative normal velocity changes its sign and is reduced 
by a factor $1-r$. 
Energy dissipation is proportional to $\diss=1-r^2$, so that
the elastic limit $r=1$ implies $\diss=0$, i.\,e.\ no dissipation, 
while $r < 1$ implies $\diss > 0$.  

\subsection{Event-Driven Molecular Dynamics}
\label{sec:ed}

The simulation of hard spheres can be handled efficiently 
with event-driven molecular dynamics \cite{lubachevsky91,sm:par}. 
The collisions are the events which have to be treated by 
the algorithm.
Between these collisions the particles move on trivial trajectories 
and so the algorithm can easily compute the point of time $t_{12}$ 
of the next collision of two particles 1 and 2 as
\begin{multline}
  t_{12} = t_0 - \vect{r}_{12} \cdot \vect{v}_{12} / v_{12}^2  \\
    +\sqrt{(\vect{r}_{12} \cdot \vect{v}_{12})^2
      - \left( r_{12}^2-d^2 \right) v_{12}^2} 
  \; / v_{12}^2 \;,
\end{multline}
where $\vect{v}_{12}=\vect{v}_2(t_0)-\vect{v}_1(t_0)$ and 
$\vect{r}_{12}=\vect{r}_2(t_0)-\vect{r}_1(t_0)$ are the relative velocities and
positions of the particles at time $t_0$, 
and $d$ is the diameter of a particle.

The algorithm processes the events one after another.
After a collision the positions and velocities of the two involved particles 
are updated; the state of all other particles remains unchanged.
For the two colliding particles, new events are calculated and the next
future event is stored in the event priority queue 
for both particles. 
The next event is obtained from the priority queue, 
the new positions and velocities after the collision for the collision partners
are updated, and so on.
Neighborhood search is enhanced with standard linked cell methods 
\cite{allen87}, where the cell change of a particle is treated 
as a new event type.
The details of the algorithm can be found in 
\cite{lubachevsky91,lubachevsky92,sm:par}.

\subsection{Avoiding the Inelastic Collapse with the TC Model}
\label{sec:tc}

Our model makes use of hard spheres 
with an infinitely stiff interaction potential.
This is an idealization of real physical particles and
can lead to the dramatic consequence of inelastic collapse: 
An infinite number of collisions occurs in finite time.
This singularity is unphysical, of course, and a major drawback 
for numerical simulations, too.
But it has been shown that one can circumvent this artefact of the model 
in the following way \cite{luding98f}:
If two consecutive collisions of a particle happen within 
a small time $t_c$, dissipation is switched off 
for the second collision.
This timestep $t_c$ corresponds to the duration 
of the contact of physical particles.

There exist other deterministic and random models which prevent 
inelastic collapse. Even though many of them lack a solid theoretical 
background and physical motivation,
their details should be insignificant for the physical evolution 
of the system anyway, since only a small negligible fraction 
of the particles in the system is involved in the inelastic collapse. 
For an extensive discussion see \cite{luding98f}.

\section{Numerical Experiments}
\label{sec:numeric}

The simulation is started from a homogeneous system, prepared as described in 
\sref{init}.
Depending on the dissipation $\diss$, the density $\dens$, 
and the number $N$ of particles, the system remains in 
the homogeneous cooling state (HCS) for some time, 
until clustering starts and the system becomes inhomogeneous.

We discuss the evolution of the kinetic energy and
the collision frequency of the system in \sref{energy}.
In \sref{cluster} we investigate the clustering 
by means of appropriate measures of the cluster size distribution.
In \sref{exponent} and \ssref{consequences} 
we encounter interesting parallels with percolation theory
and discuss several critical exponents. 

\subsection{Kinetic Energy and Collision Frequency}
\label{sec:energy}

\begin{figure*}
  \begin{center}
    \includegraphics*[width=\hgxlen]{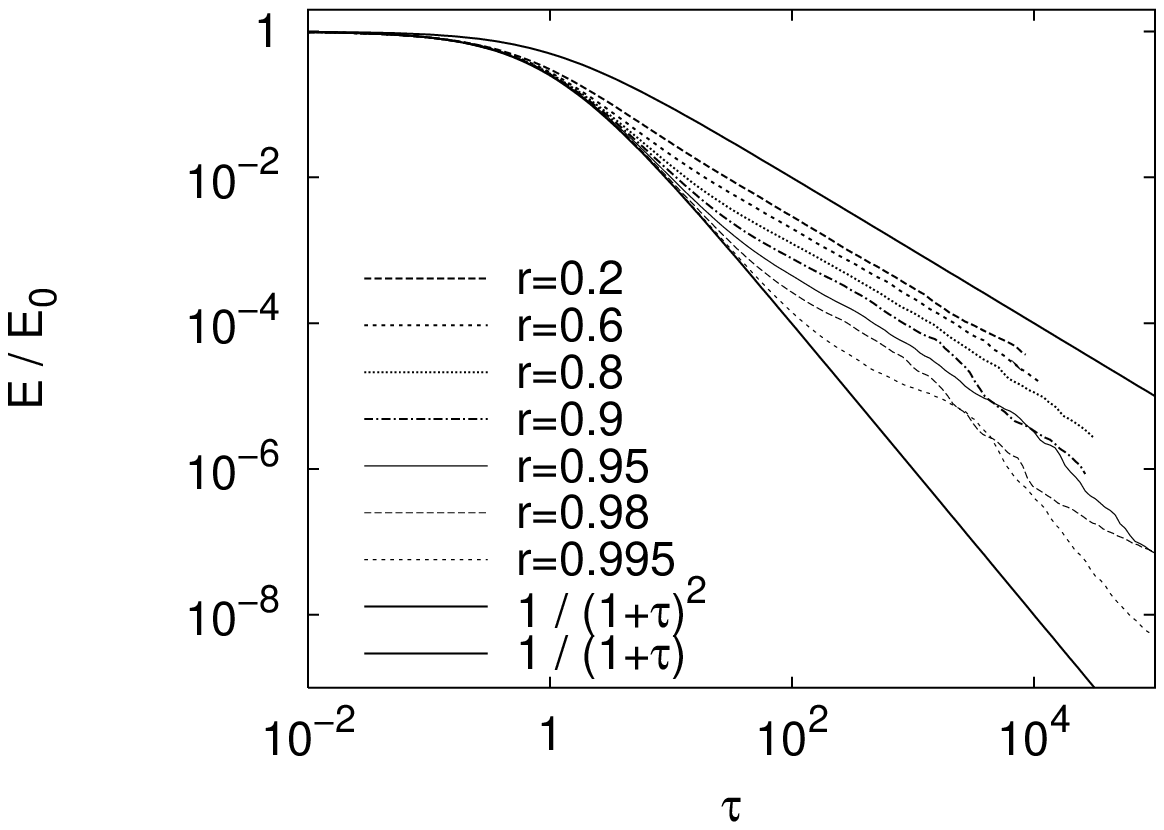}\hfill
    \includegraphics*[width=\hgxlen]{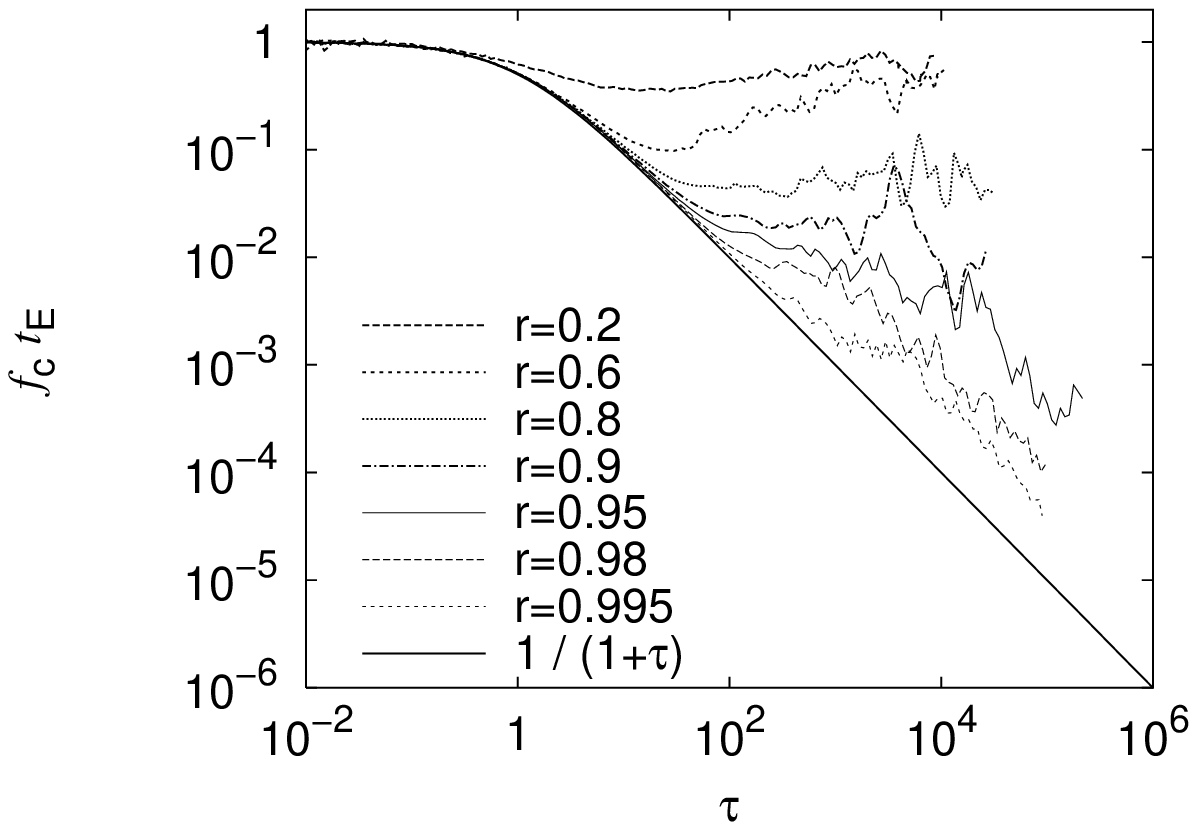}\\*
  \end{center}
  \caption{Decay of the kinetic energy $E$ (left) 
    and the collision frequency $f_c$ (right) plotted against scaled
    time $\tau$ in a 2D system with $N=316^2=99856$ particles, 
    volume fraction $\dens=0.25$, 
    and different restitution coefficients $r$ 
    (increasing $r$ from top to bottom). 
    The thick solid lines correspond to the 
    theoretical predictions as given in the inset.
    }
  \label{fig:energy2d}
\end{figure*}

\begin{figure*}
  \begin{center}
    \includegraphics*[width=\hgxlen]{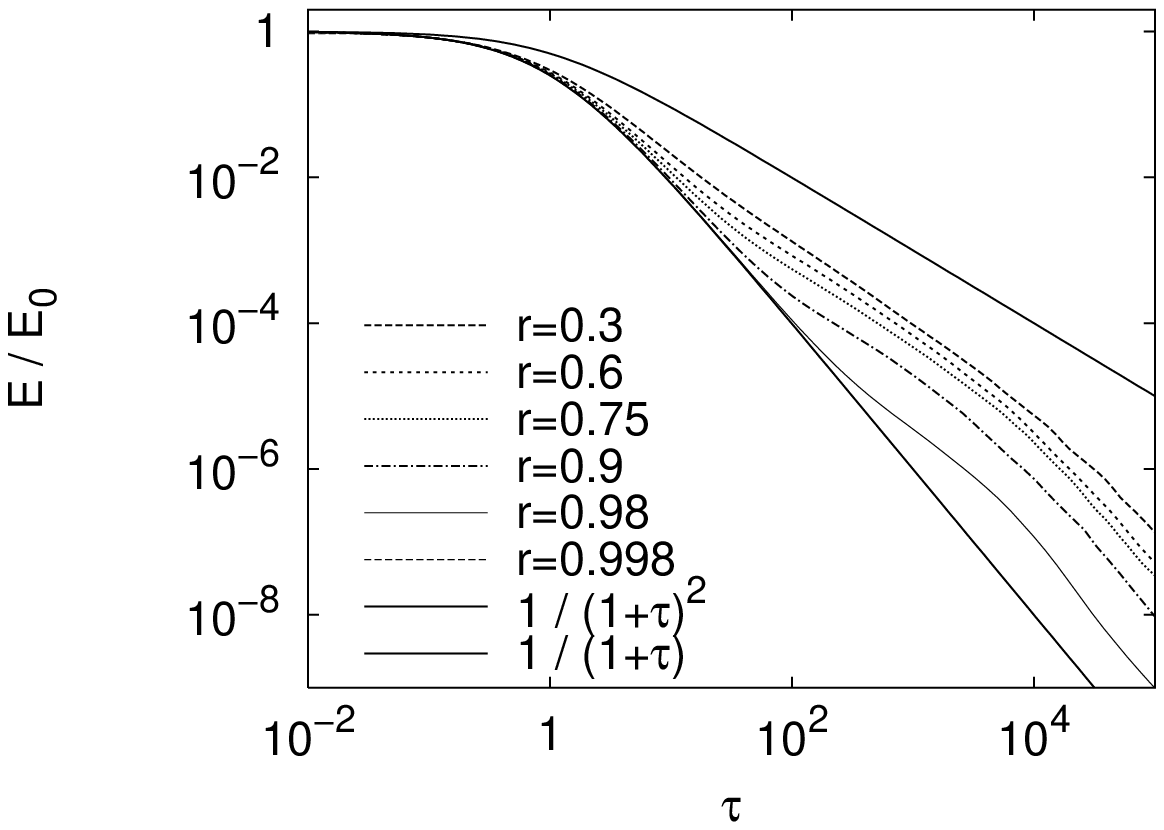}\hfill
    \includegraphics*[width=\hgxlen]{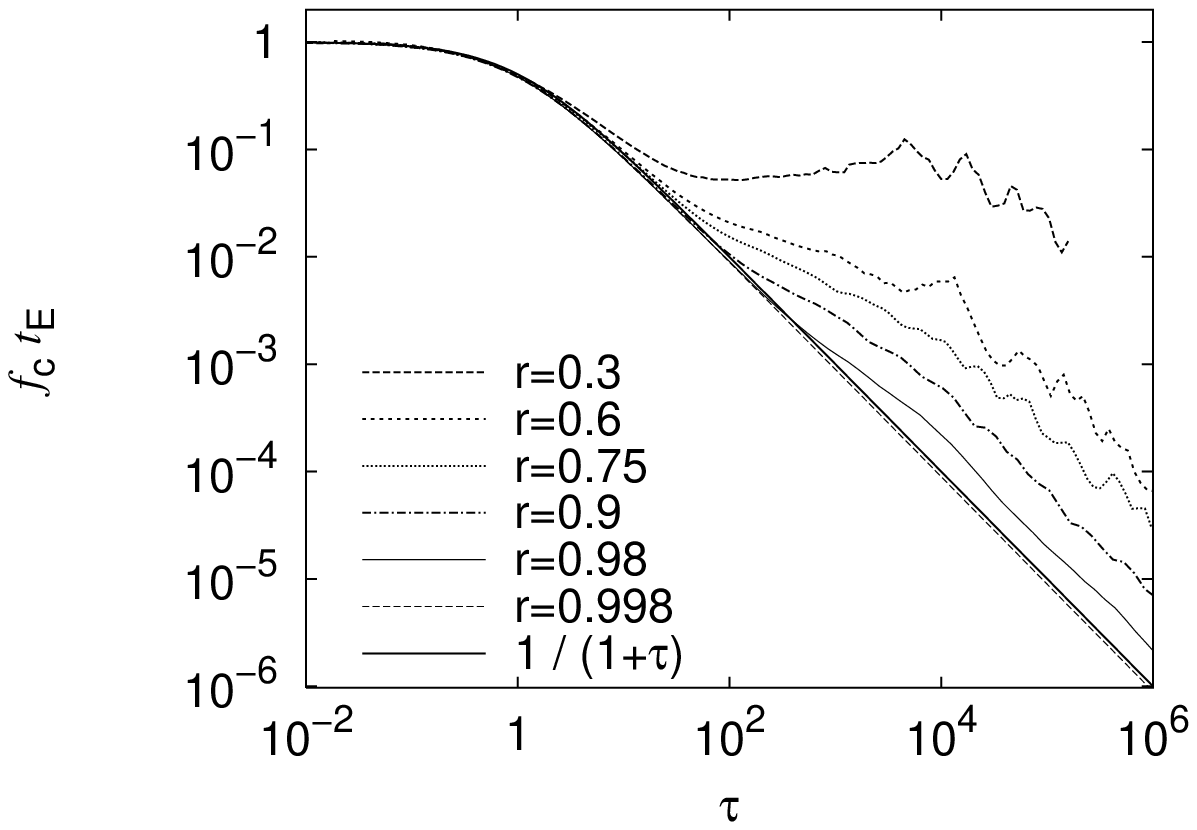}\\*
  \end{center}
  \caption{Decay of the kinetic energy $E$ (left)
    and the collision frequency $f_c$ (right) plotted against scaled
    time $\tau$ in a 3D system with $N=80^3=512000$ particles, 
    volume fraction $\dens=0.25$, 
    and different restitution coefficients $r$ 
    (increasing $r$ from top to bottom). 
    The thick solid lines correspond to the 
    theoretical predictions as given in the inset.
    }
  \label{fig:energy3d}
\end{figure*}

Dissipative collisions lead to a decay of the kinetic energy 
and the collision frequency
(see \fref{energy2d} and \fref{energy3d}).
From these figures three different regimes can be clearly distinguished.
They are: 

(1) The {\em homogenous cooling state} (HCS) at the beginning, 
when no clusters have formed yet, is well understood \cite{haff83,luding98d}.
The decay of the kinetic energy $E$ is governed by the equation
\begin{align}
  \label{eq:energy}
  E(\tau) &= \frac{E(0)}{(1 + \tau)^2} \;,
\end{align}
with the scaled time $\tau= \frac{\diss}{2 D} \frac{t}{t_E}$.
$D$ is the dimension of the system and $t_E$ is 
the initial Enskog collision time
$t_E=(\sqrt{\pi} d)/(2 \sqrt{D} v \dens g_d(\dens))$.
$v$ is the mean velocity of a particle, $d$ its diameter, 
$\dens$ is the volume fraction, and $g_d$ is the contact probability.
In 2D $g_d(\dens)=(1-7\dens/16)/(1-\dens)^2$ and in 3D 
$g_d(\dens)=(1-\dens/2)/(1-\dens)^3$.

The evolution of the collision frequency per particle $f_c(\tau)$ 
with time is given by
\begin{align}
  \label{eq:collision}
  f_c(\tau) &= t_E^{-1}(0) \sqrt{\frac{E(\tau)}{E(0)}} \;
\end{align}
and it is the natural time scale controlling the evolution of 
the system in the HCS.

\begin{figure*}
  \begin{center}
    \includegraphics*[width=\hgxlen]{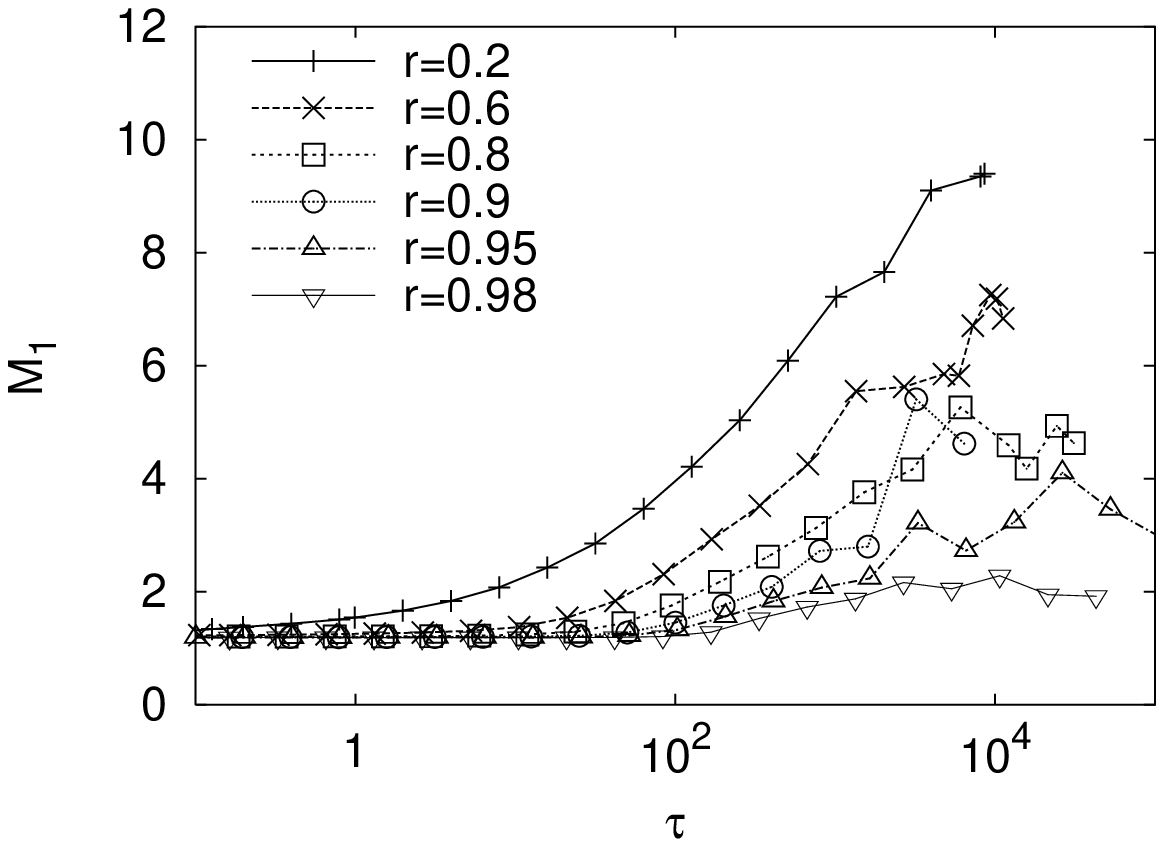}\hfill
    \includegraphics*[width=\hgxlen]{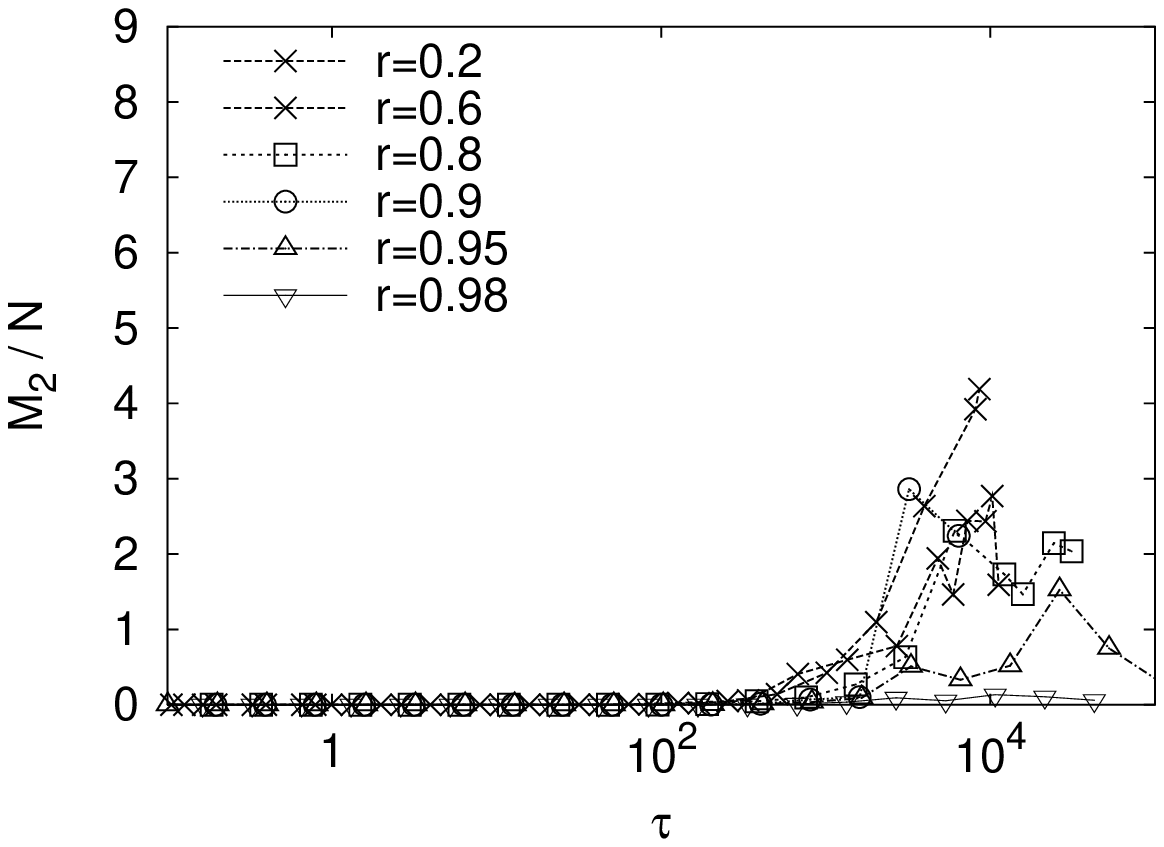}\\*
  \end{center}
  \caption{Growth of the 1st moment $M_1$ (left) 
    and the 2nd moment $M_2$ (right)
    of the cluster size distribution plotted against scaled
    time $\tau$ in a 2D system with $N=99856$ particles, 
    volume fracion $\dens=0.25$, 
    and different restitution coefficients $r$ 
    (increasing $r$ from top to bottom).
    }
  \label{fig:mom12_2d}
\end{figure*}

\begin{figure*}
  \begin{center}
    \includegraphics*[width=\hgxlen]{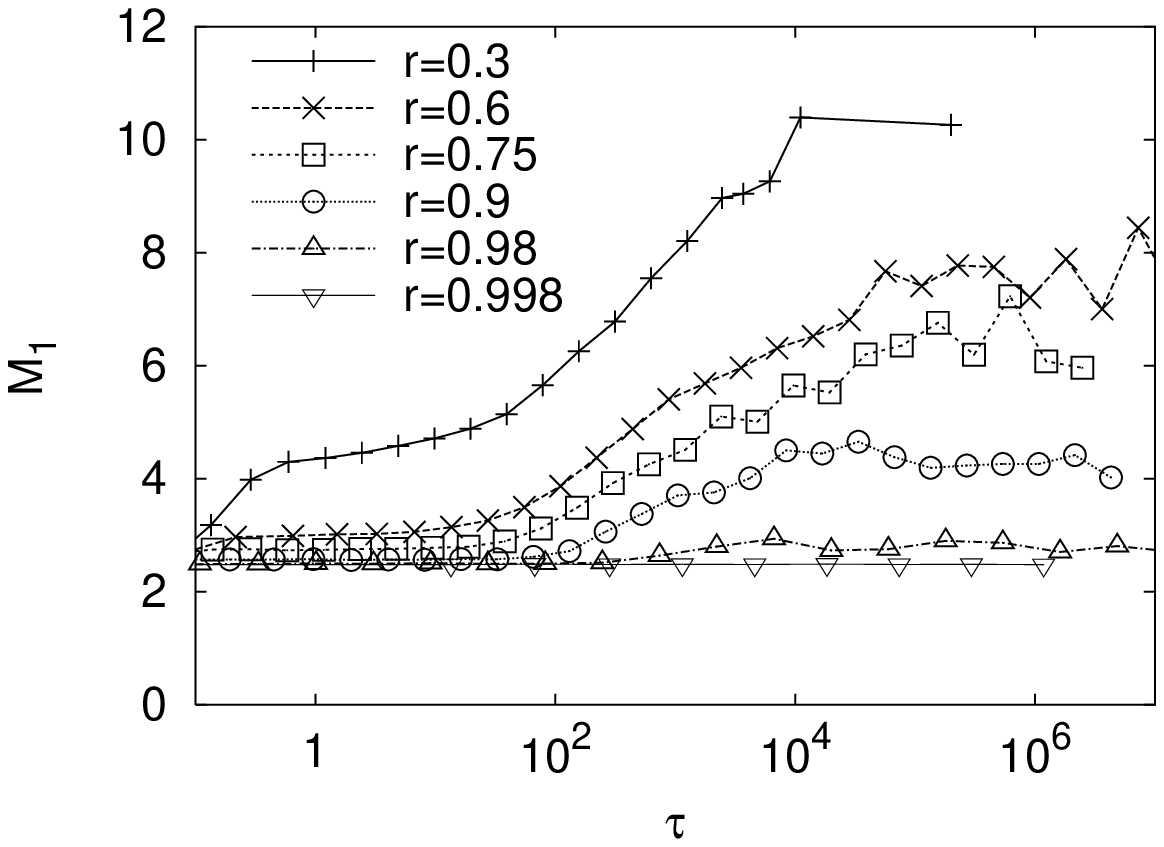}\hfill
    \includegraphics*[width=\hgxlen]{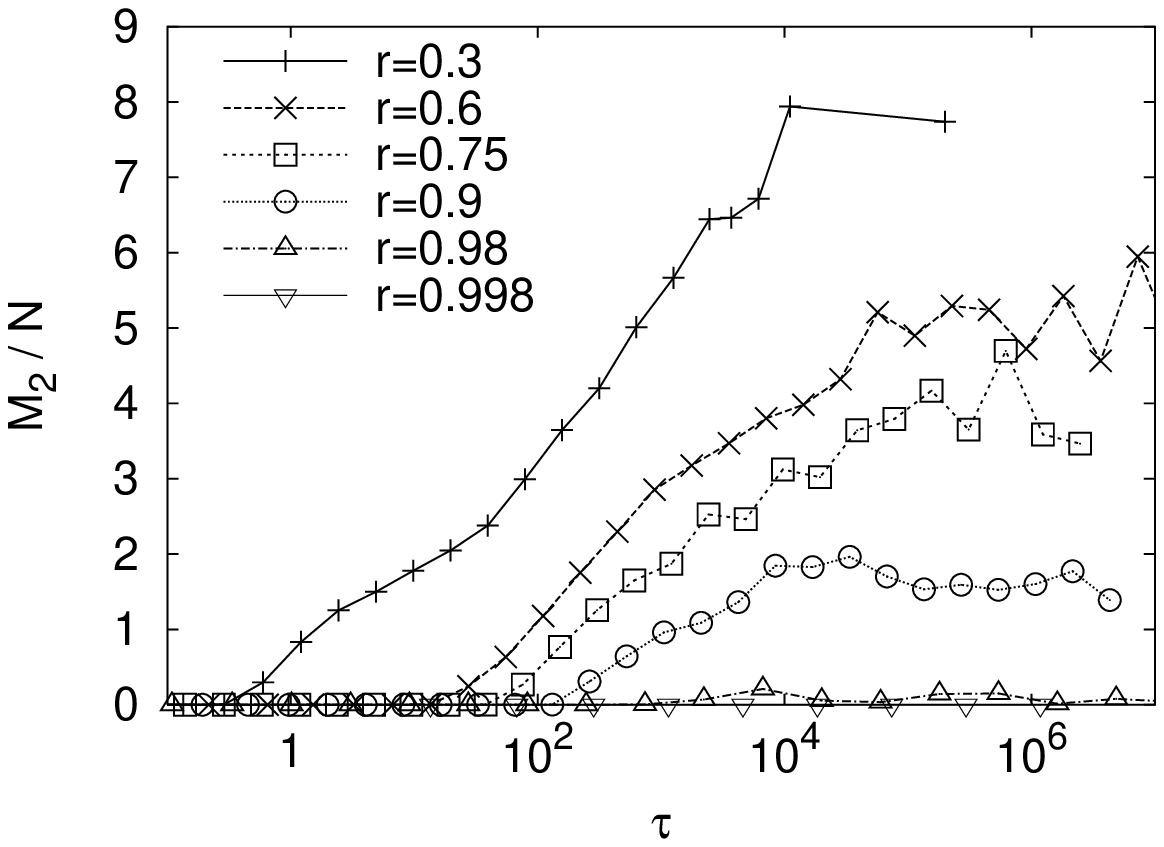}\\*
  \end{center}
  \caption{Growth of the 1st moment $M_1$ (left) 
    and the 2nd moment $M_2$ (right) of the cluster size distribution 
    plotted against scaled time $\tau$ 
    in a 3D system with $N=512000$ particles, volume fracion $\dens=0.25$, 
    and different restitution coefficients $r$ 
    (increasing $r$ from top to bottom).
    }
  \label{fig:mom12_3d}
\end{figure*}

(2) When clusters start to grow, the decays deviate from these laws.
In the {\em cluster growth regime}, the decay of the kinetic energy slows down.
Furthermore, the collision frequency fluctuates a lot 
and can even increase.  Note that the collision frequency
is no longer a natural time scale of the system, since it is 
mainly determined by cluster-cluster collisions, where it 
increases strongly.
The deviation from \eref{energy} occurs earlier and is more dramatic 
for larger dissipation $\diss$, i.\,e.\ smaller $r$.
However, the cluster growth regime is characterized 
by an energy decay around $E \sim \tau^{-1}$, independently of $r$, 
cf.~\cite{nie02}.
(In 3D the exponent might be slightly larger than unity; a best fit yields
$E \sim \tau^{-1.1 \pm 0.1}$.) 
In contrast, the evolution of the collision frequency with time
depends on $r$ during cluster growth.
This regime is more distinct for large dissipation $\diss$ and 
large system sizes $N^{1/D}$ and can e.\,g.\ barely be seen for $r=0.98$ in
\fref{energy3d}. 

(3) Finally, when the largest cluster in the population has reached 
system size in the {/em saturation regime}, 
the cooling resembles the homogeneous cooling state 
in so far that $E(\tau) \propto \tau^{-2}$ and $f_c(\tau) \propto \tau^{-1}$,
even if the latter still shows large fluctuations.
In \fref{energy2d} this regime is not clearly visible, because 
the simulation times are not long enough for this system. 

More details of the 2D situation were discussed 
in \cite{mcnamara96,luding98f} and references therein.
Further studies on the three-dimensional systems are in progress.

\subsection{Cluster Growth}
\label{sec:cluster}

The energy loss of the particles first leads to a reduced separation velocity 
after collision and eventually to the formation of clusters.
But the definition of a cluster suffers from the fact that 
it takes an infinite number of collisions 
for the particles to stay in permanent contact with each other 
\cite{luding99}.
So we use the following (geometrical) definition:
Two particles belong to the same cluster, if their distance 
is smaller than $s=0.1$ particle diameters.
The choice of $s$ is arbitrary and shifts the results;
the qualitative behavior of the quantities discussed below 
does not depend on $s$, as long as it is neither too small nor too large
\footnote{
For a detailed study of different $s$ in 2D see \cite{luding99};
in 3D, the $s$ dependence of some quantities seems to be stronger
than for others.}.

The moments $M_k$ of the cluster size distribution are defined as
\begin{align}
  M_k &:= \frac{1}{n_c} \sum_i{ i^k n_i } \, ,
\end{align}
where $n_c$ denotes the total number of clusters and 
$n_i$ the number of clusters of size $i$.

In many cases there are a lot of small clusters and one large cluster of size 
$N_x$, which contains a macroscopic fraction $m_x:=N_x/N$ 
of the total number $N$ of particles.
Therefore, we also define reduced moments $M'_k$,
which do not include the largest cluster. 

In \fref{mom12_2d} and \fref{mom12_3d} the growth of the clusters 
can be seen 
on the basis of the first and second moments.
After several collisions, particles start to cluster and the moments of the 
cluster size distribution grow until they reach their ``saturation'' value
\footnote{Saturation means that there is a transition to a regime with a 
much slower change of the moments. The dynamics is not finished, however,
since the large cluster still can collect particles or break into pieces.}.
A numerical analysis reveals that the increase in $M_1$ and $M_2$ is mainly 
due to one large cluster which grows until it reaches its maximum size.
In this final state this cluster contains a macroscopic fraction $m_x$ 
of the particles (see \fref{momn}).

\begin{figure}[htbp]
  \begin{center}
    \includegraphics*[width=\gxlen]{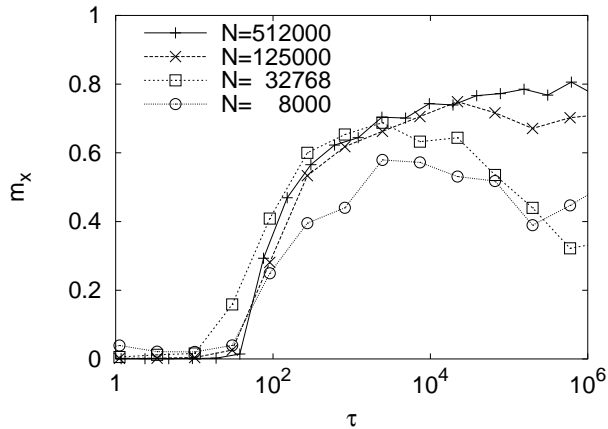}\\*
  \end{center}
    \caption{Growth of the large cluster $m_x$ in a 3D system with 
      $N$ particles, volume fraction $\dens=0.25$, 
      and a restitution coefficient $r=0.75$.}
    \label{fig:momn}
\end{figure}

The onset time of cluster growth and also the final size $m_x$ 
of the large cluster depend strongly on the restitution coefficient $r$
(see \fref{mom12_2d} and \fref{mom12_3d}).
At low dissipation rates, for a long time nothing interesting happens 
and finally small and strongly fluctuating clusters appear.
High dissipation leads to almost immediate cluster growth 
and a very large cluster at last.
On the other hand, the system size $N$ does not seem to affect 
the behavior of the system provided that $N$ is not too small 
(see \fref{momn}).

It has been shown \cite{mcnamara96,sela95,goldhirsch93}, that 
for infinite system size the system is always unstable 
to the formation of clusters, whereas for a finite system size 
the density $\dens$ and the dissipation $\diss=1-r^2$ must not be too small.
With our choice $N \ge 10^5$ and $\dens=0.25$ we expect no cluster formation 
below the critical dissipation $\diss_c=10^{-4}$ in our 2D system 
and $\diss_c=10^{-2}$ in our 3D system \cite{mcnamara96}.
This is in good agreement with our results for the 3D system, where 
cluster formation still can be seen with dissipation $\diss=0.04$, 
but not for $\diss=0.004$ (see \fref{mom12_3d}).

\begin{figure*}
  \begin{center}
    \includegraphics*[width=\hgxlen]{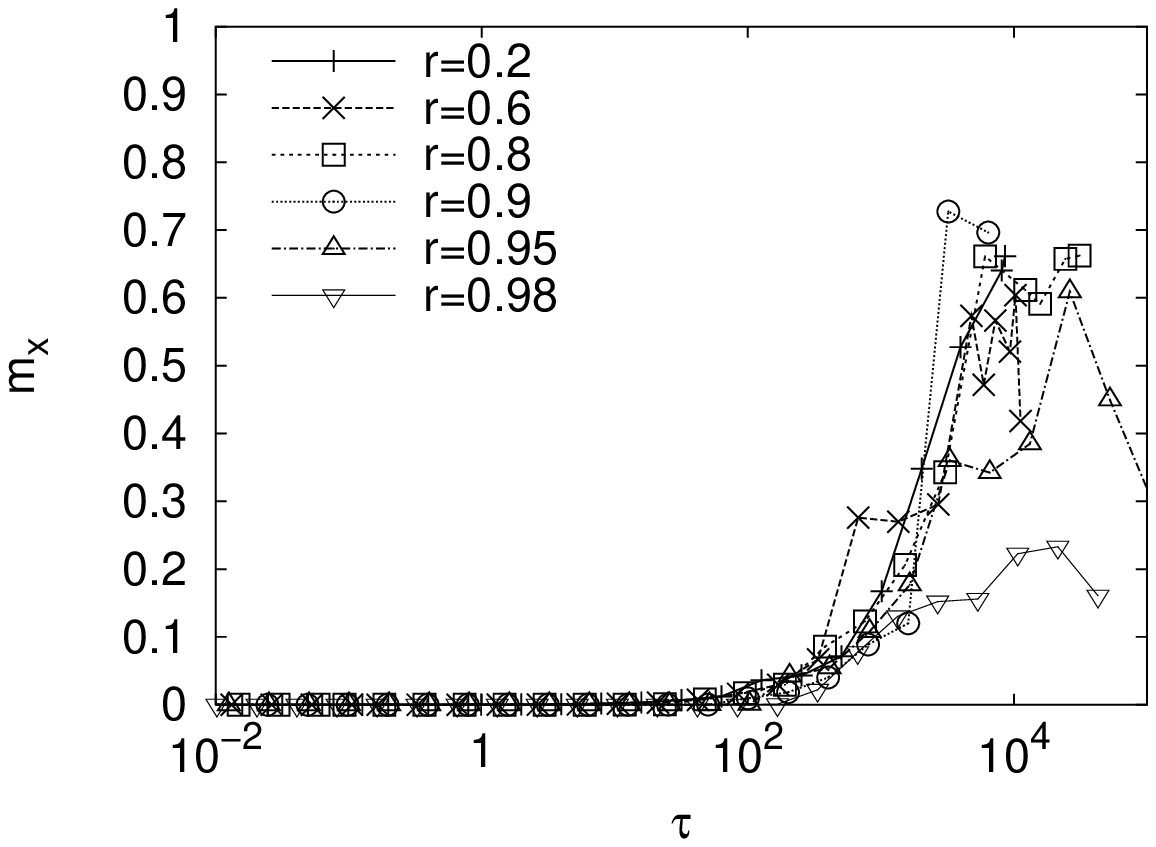}\hfill
    \includegraphics*[width=\hgxlen]{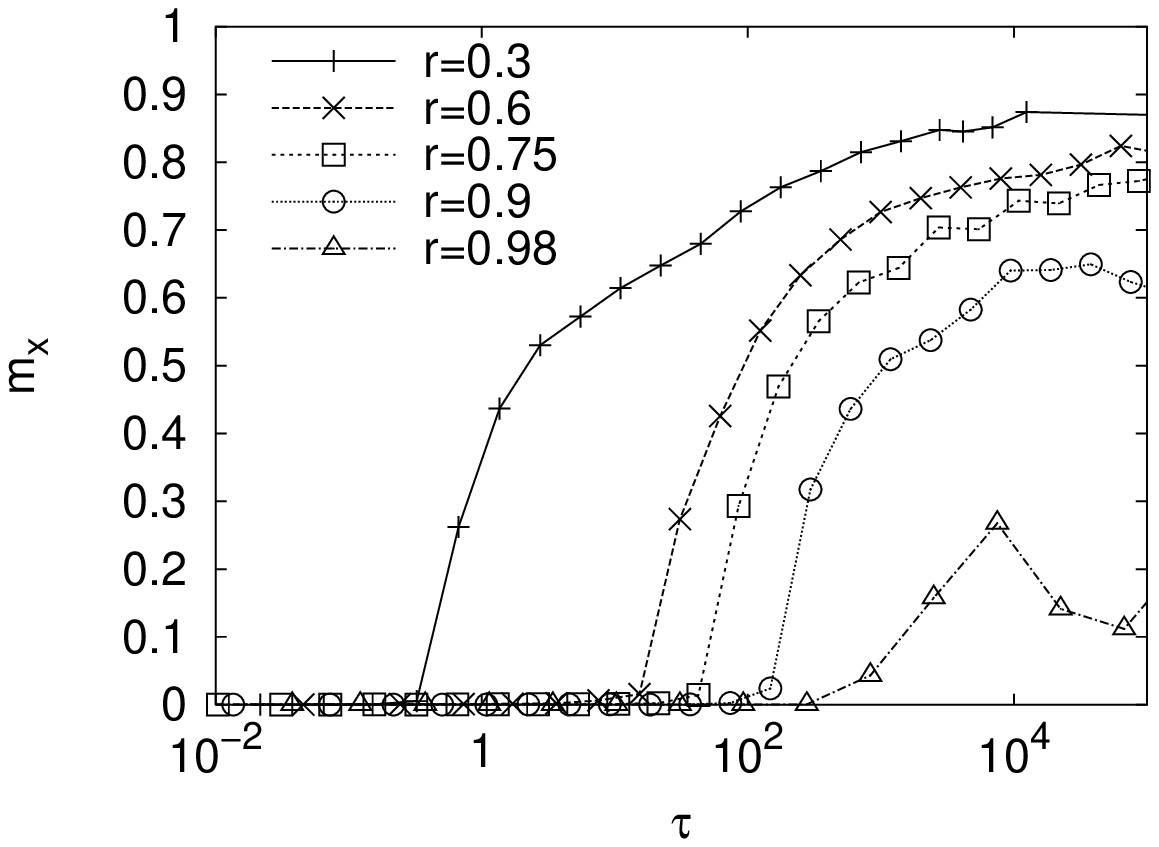}\\*
  \end{center}
    \caption{Growth of the large cluster $m_x$ in 2D (left) and 3D (right)
      for different restitution coefficients $r$.
      }
    \label{fig:momx}
\end{figure*}

\begin{figure*}
  \begin{center}
    \includegraphics*[width=\hgxlen]{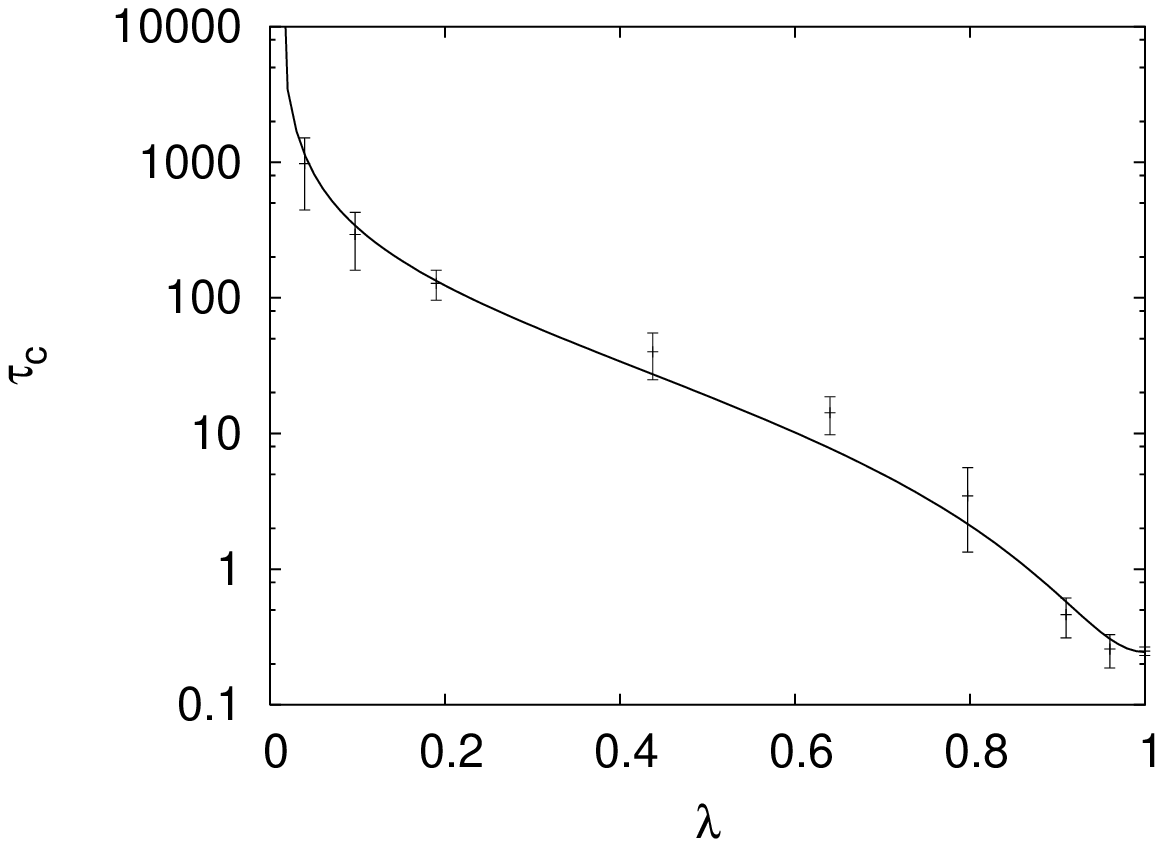}\hfill
    \includegraphics*[width=\hgxlen]{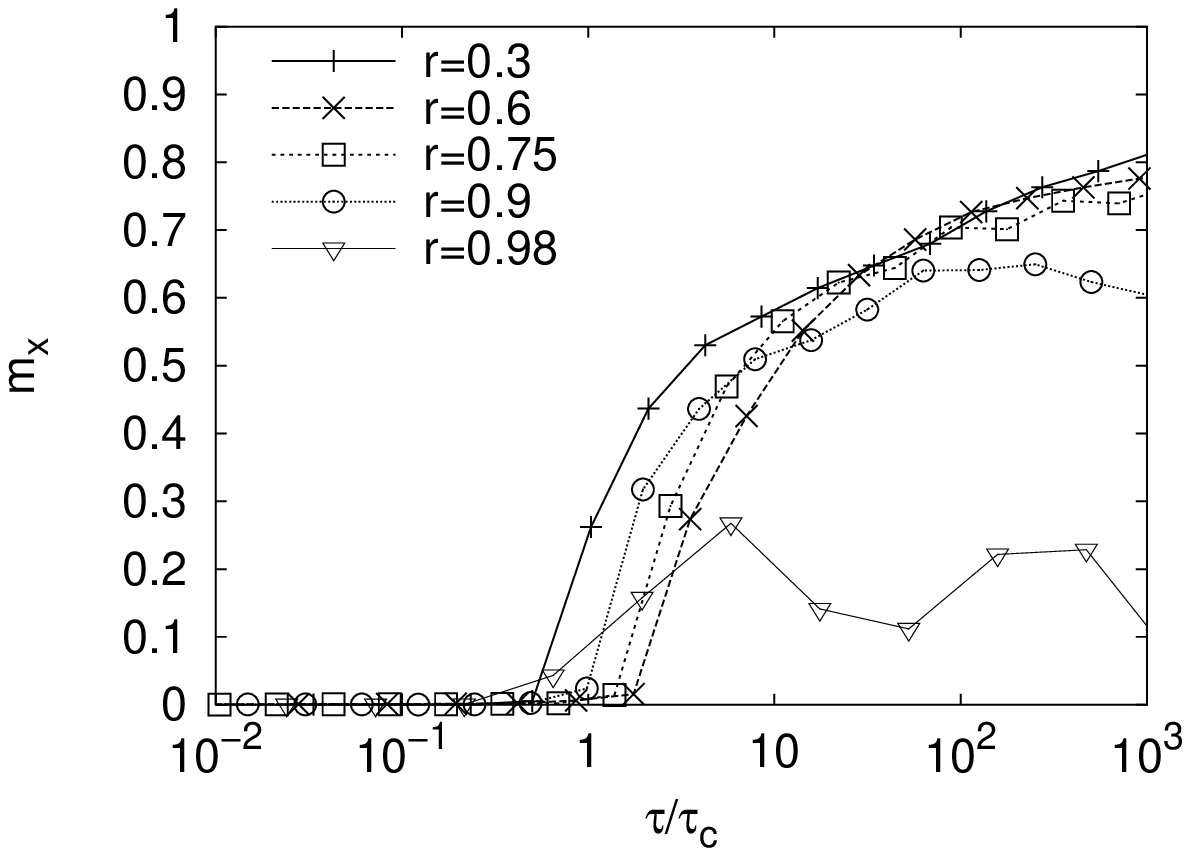}\\*
  \end{center}
    \caption{
      (left) Scaling behavior of the onset $\tg$ of the cluster growth
      in 3D depending on the dissipation rate $\diss$. 
      The curve is \eref{scgamma}.
      \\*
      (right) With rescaled time $\tau / \tg$ according to 
      \eref{scgamma} the curves of the growth of the large cluster in 3D 
      almost collapse.
      }
    \label{fig:scgamma}
\end{figure*}

As can be seen in \fref{momx} (left) the onset $\tg$ 
of the growth of the large cluster in 2D does not depend 
on the restitution coefficient $r$ explicitly. 
(Implicitly, $\tg$ is dependent on $r$ via rescaled time, of course.)
In contrast $\tg$ does strongly depend on $r$ in a 3D system 
(see \fref{momx} (right)).

An empirical formula, which gives the dependency rather accurately
over orders of magnitude in $\diss$, is
\begin{align}
  \tg(\diss) &= \tau_1 + c \, \frac{(1-\diss)^2}{\diss-\diss_c}  \;,
  \label{eq:scgamma}
\end{align}
where $\tau_1=0.24 \pm 0.02$ and $c=36 \pm 5$ are fit parameters (for
$s=0.1$). 

In the limit $\diss \to 1$, one has $\tg \to \tau_1$, where
$\tau_1$ is proportional to the rescaled collision frequency and
corresponds to $t=2 D \tau_1 t_E / \diss \approx 1.5 \, t_E$.
This constant stems from the fact that the particles need at least
one collision to cluster, however, it also depends on $s$.  
In the limit $\diss \to \diss_c$ the time $\tg$ 
diverges: the particles never start to cluster
\footnote{
Since we focused on strong dissipation in this study,
rather than the limit $\diss \to \diss_c$, we can
not draw a conclusion on the functional behavior from our
limited data.}.

of magnitude in
Rescaling time in \fref{momx} (right) according to \eref{scgamma} 
results in \fref{scgamma} (right). 
There, the onset of the clustering happens approximately at the same
rescaled time, whereas these times differ by 4 orders of magnitude in
\fref{momx} (right).
The fluctuations in $\tau_c$ are not systematically
dependent on $r$.

Another difference between the 2D and 3D system in \fref{momx} is the 
fact that the cluster growth in 2D starts very smoothly, 
whereas the beginning of the clustering is very sharp in 3D.

\subsection{Critical exponents in 3D}
\label{sec:exponent}

\begin{figure*}
  \begin{center}
    \includegraphics*[width=\hgxlen]{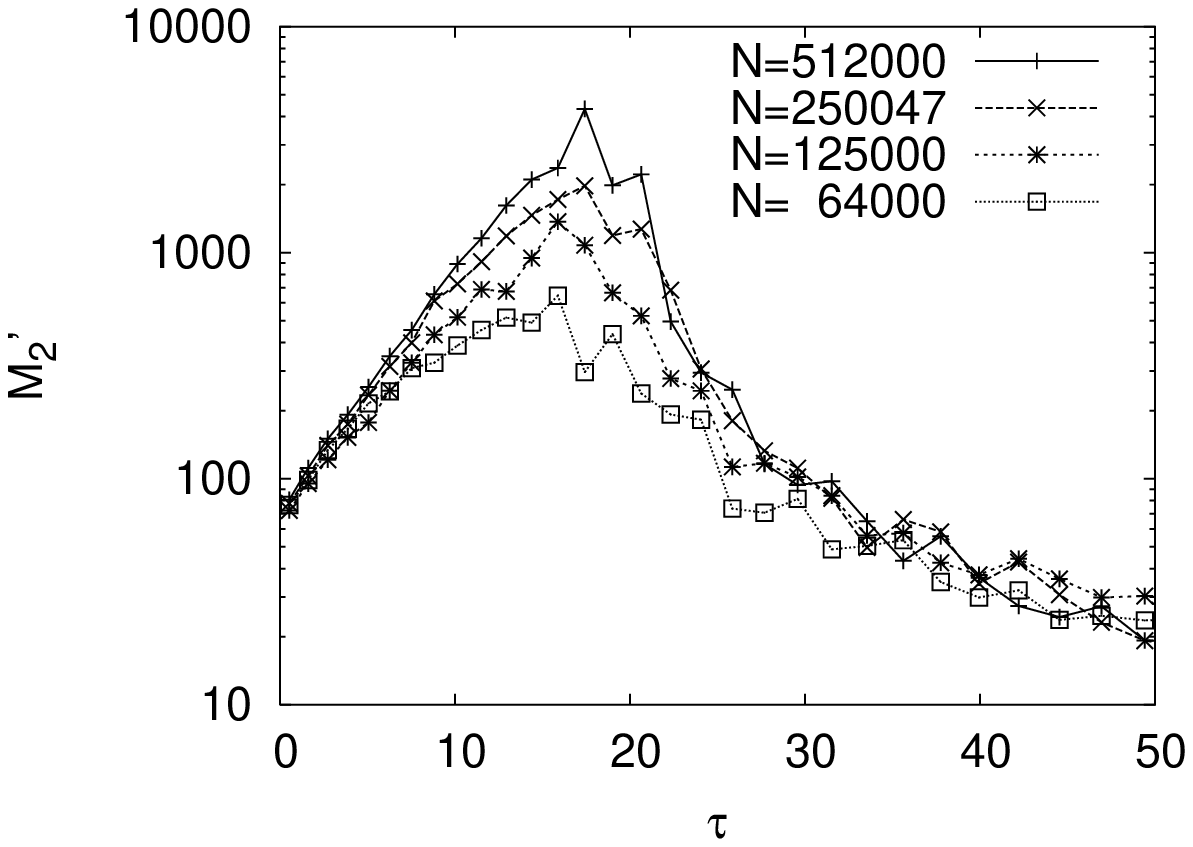}\hfill
    \includegraphics*[width=\hgxlen]{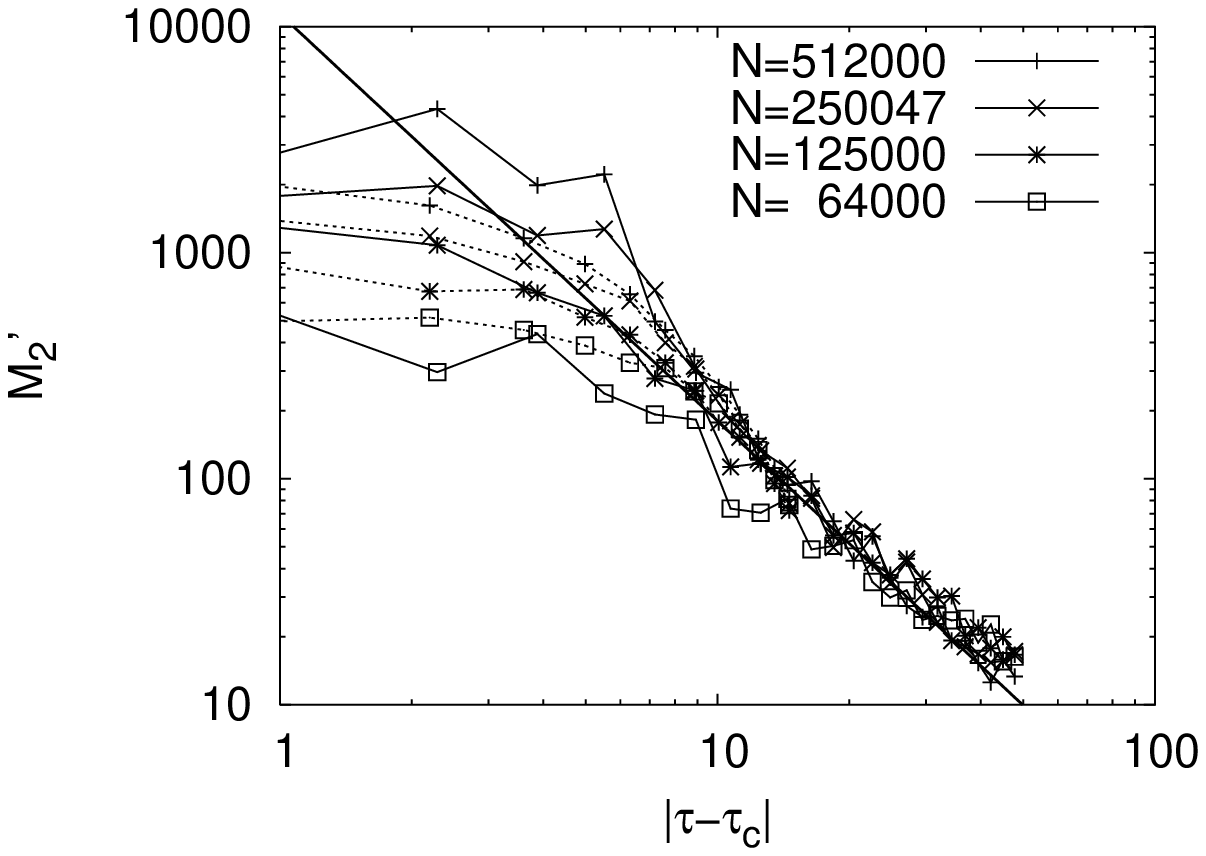}\\*
  \end{center}
    \caption{Reduced 2nd moment $M'_2$ of the cluster size distribution 
      in 3D systems with different $N$, 
      volume fraction  $\dens=0.25$, 
      and a restitution coefficient  $r=0.6$.
      The data are averaged over 10-20 different simulation runs.\\
      (left) The curves show a peak at the onset $\tg$ of the cluster growth.\\
      (right) Scaling behavior of $M'_2(\tau)$ against
      $|\tau-\tg|$; each data curve has two branches, 
      solid and dotted lines connect data points for $\tau>\tg$ and 
      $\tau<\tg$, respectively. 
      The additional line has a slope of -1.8.
    }
    \label{fig:peak}
    \label{fig:peakexp}
\end{figure*}

As the moments of the cluster size distribution are clearly dominated
by the large cluster, we will now study the reduced second moment $M'_2$, 
which does not include the large cluster.
\Fref{peak} shows that $M'_2$ is small most of the time, 
except for a peak at $\tg$.
This gives us a clean definition of $\tg$.
But what is more interesting, parallels to percolation theory arise.
In percolation theory \cite{stauffer94} one studies 
the scaling behavior of certain quantities 
depending on the occupation probability $p$ around 
the percolation threshold $p_c$.
One result for the reduced second moment $M'_2(p)$ is, e.\,g., 
\begin{align}
  M'_2(p) \sim \abs{p-p_c}^{-\gamma} \;,
  \label{eq:pgamma}
\end{align}
where $\gamma$ is a universal critical exponent.
In order to transfer this result to cluster growth, we replace $p$ 
with $\tau$ and $p_c$ with $\tg$:
\begin{align}
  M'_2(\tau) \sim \abs{\tau-\tg}^{-\gamma}
  \label{eq:taugamma}
\end{align}
Indeed we find in \fref{peakexp} the same power law behaviour.
All data for different system sizes collapse on the same master-curve and
even the exponent $\gamma = 1.8 \pm 0.1$ is in 
agreement with the 3D percolation problem result $\gamma = 1.80$
after the beginning of clustering.

Now, if we study the amplitude of this peak,
percolation theory tells us that the maximum sits at
\begin{align}
  p_{max} \sim p_c \, (1-a L^{-1/\nu}) \;,
\end{align}
where $L$ is the system size, $a$ is a constant, 
and $\nu$ is another critical exponent.
Thus we expect:
\begin{align}
  M'_2(p_{max}) \sim L^{\gamma/\nu} \;.
\end{align}
As the system size $L \sim N^{1/3}$, the maximum value of the peak 
in \fref{peak} scales as
\begin{align}
  M'_{2,max}(N) \sim N^{\gamma / 3\nu} \;.
\end{align}

\begin{figure}
  \begin{center}
    \includegraphics*[width=\gxlen]{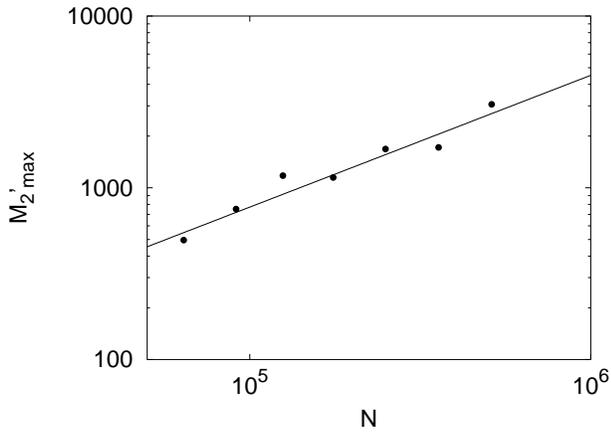}\\*
  \end{center}
    \caption{Maximal amplitude of the reduced 2nd moment $M'_2$ 
      of the cluster size distribution (see \fref{peak}).
      The line has a slope of 0.77.}
    \label{fig:peakmax}
\end{figure}
Here we can also verify the power law behavior (see \fref{peakmax}).
Our simulations yield the result $\gamma / 3\nu = 0.77 \pm 0.09$, 
which leads to $\nu=0.78 \pm 0.14$.
Within the rather large margin of errors this result is close to 
the 3D percolation problem $\nu=0.88$, too.
 
A third exponent $q$ \footnote{The common notation for this exponent is $\tau$.
In order to avoid name conflicts we call it $q$.}
is given by the cluster size population at time $\tg$:
\begin{align}
  n_i (\tg) \sim i^{-q}\;,
\end{align}
where $n_i$ is the number of clusters of size $i$.
\begin{figure}
  \begin{center}
    \includegraphics*[width=\gxlen]{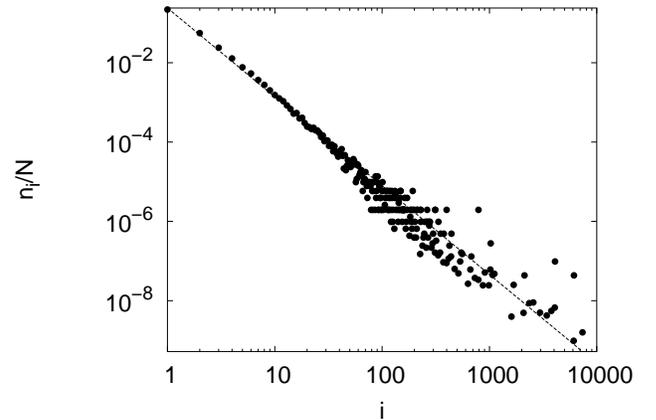}\\*
  \end{center}
    \caption{
      The points give the numbers $n_i$ of clusters of size $i$ at 
      time $\tg$ in a 3D simulation with $N=512000$ particles 
      and a restitution coefficient $r=0.75$.
      The line has a slope of -2.2.
      }
    \label{fig:sizedistribution}
\end{figure}
\fref{sizedistribution} yields $q=2.2 \pm 0.2$.
This result is in agreement with the prediction of percolation 
theory $q=2.2$, too. 

Is this similarity of clustering in 3D and percolation coincidence 
or is there a deeper reason?

\subsection{Consequences}
\label{sec:consequences}

Percolation theory makes many universal predictions for
randomly disordered systems.
The particles in a granular gas form such a disordered system.
There have been other attempts to apply percolation theory 
to granular systems \cite{tobochnik99}, and also more advanced
phase ordering models have been used to parallel the clustering
dynamics \cite{das03a,das03b}.
Our results indicate that such attemps can be justified, even though
the percolation problem is purely static, whereas the clustering
of granular gases also involves the dynamics, i.\,e.\ momentum and
energy.  

The only thing that might seem strange at first sight is the derivation of
\eref{taugamma}.
Why can we replace the purely static quantity $p$ with time $\tau$?

In order to introduce the occupation probability $p$ 
in the clustering problem, we define it as
\begin{align}
  p(\tau):=\frac{M_1(\tau)-M_1(\tau^\star)}{M_1(\tau)}\;
\end{align}
where $\tau^\star \ll \tg$ is a time shortly after the first few 
collisions.
Now, we examine $p(\tau)$ around $\tg$.
\begin{figure*}[htb]
  \begin{center}
    \includegraphics*[width=\hgxlen]{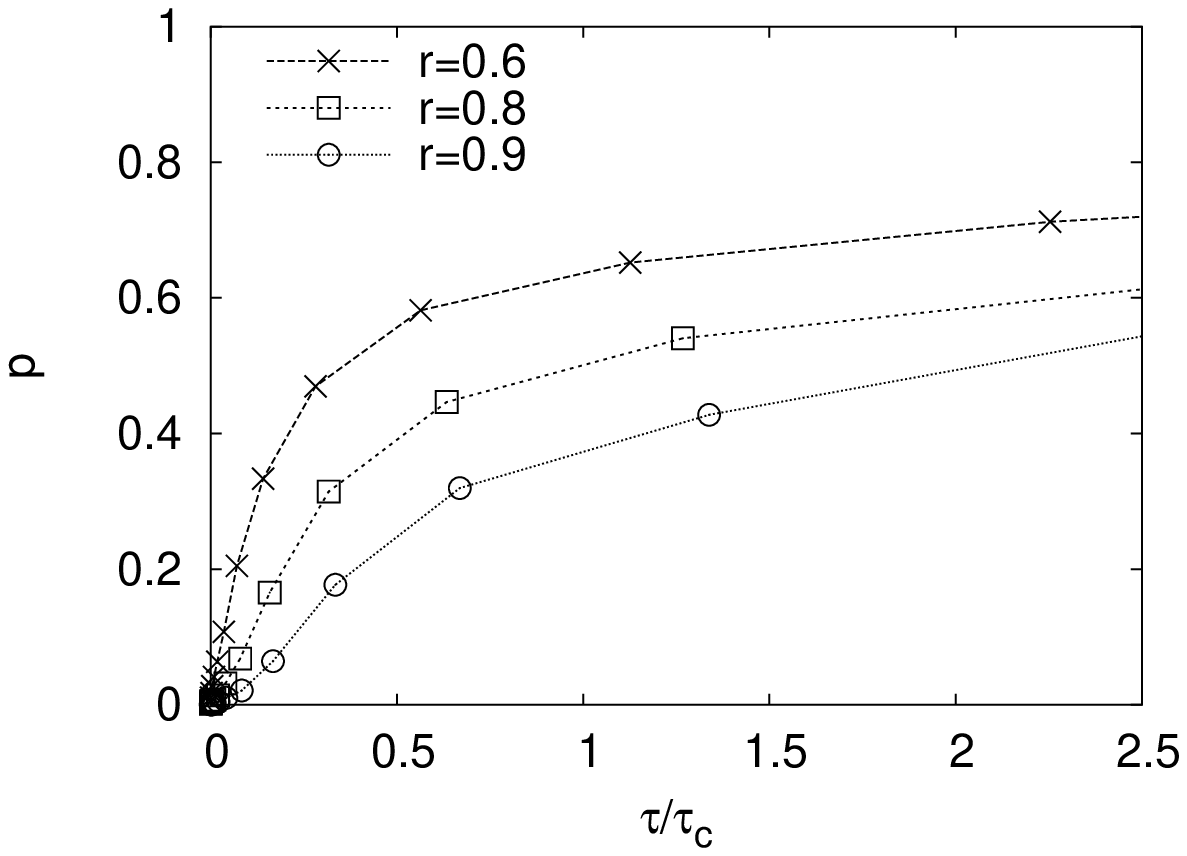}\hfill
    \includegraphics*[width=\hgxlen]{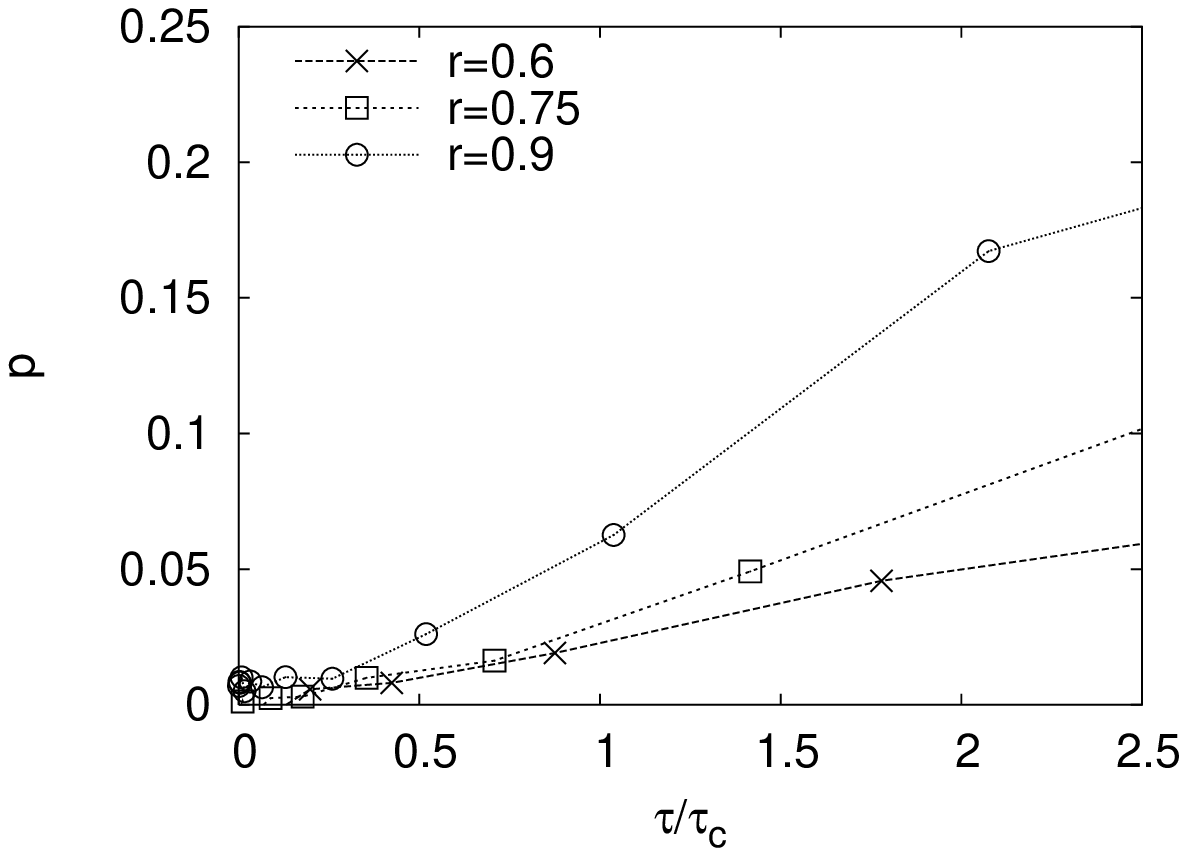}\\*
  \end{center}
    \caption{\lqq{}Occupation probability\rqq{} $p$ against
      scaled time $\tau/\tg$.\\
      (left) In 2D the data curves are highly non-linear.\\
      (right) In 3D the data curves are almost linear.
    }
    \label{fig:occprob}
\end{figure*}
\Fref{occprob} (right) shows that in 3D, $p$ is almost proportional to $\tau$.
If we make use of this linear relation and insert it in \eref{pgamma},
we arrive at the postulated formula \eref{taugamma}. 

In contrast, in our 2D system the growth of the large cluster 
happens at a rather large time $\tg$.
(Note, that $\tg$ is not very clearly defined in the 2D case, because the onset
of the cluster growth is very smooth.) 
In this late stage, the relation between $p$ and $\tau$ is highly non-linear, 
see \fref{occprob} (left).
Thus we could not find universal critical exponents for $\tau$ there.
However, it might be possible to find a linear relation 
for a different set of parameters, e.\,g.\ for a different volume fraction, 
too. 
Whether a 2D system and a 3D system with the same parameters
(like, e.\,g., volume fraction) are in fact comparable is another 
open issue.

\section{Summary and Discussion}
\label{sec:summary}

The evolution of freely cooling granular systems can be divided 
into three regimes.
First, the system is in the {\em homogeneous cooling state} (HCS).
Then, in the {\em cluster-growth} regime, clusters begin to develop 
and grow.  Finally, in the {\em saturation regime},
the clusters merge to practically one large cluster, which grows
until it reaches system size.
Besides the macroscopic fraction of particles in the
large cluster, there are still many small clusters with interesting
statistics.

In the HCS,
the decay of the kinetic energy $E$ and the collision frequency $f_c$
can be described by the simple analytical expressions 
$E(\tau) \sim (1+\tau)^{-2}$ and $f_c(\tau) \sim (1+\tau)^{-1}$. 
The collision frequency is the natural time scale here, mainly determined 
by the density and the dissipation rate of the system.
For strong dissipation, at short range, particles already 
stay closer together after only one collision, so that the moments 
of the size distribution change rather early due to a (short-range)
change in the radial pair-distribution (data not shown).

In the {\em cluster-growth} regime, 
the collision frequency shows large fluctuations because of 
cluster-cluster collisions
and cannot be predicted during cluster growth, because it
changes erratically and discontinuously.
The energy decay is characterized by $E \sim \tau^{-1}$, where
the accuracy of the exponent is limited in 3D due to the 
comparatively short duration of the cluster growth regime.
However, 
this regime shows interesting differences between two and three dimensions.
In 3D, cluster growth can be described 
by a power law behavior with the same critical exponents 
as in percolation theory.
The onset of this cluster growth $\tg$ is very sharp and does strongly depend 
on dissipation $\diss$.
In contrast, we could not find a similar behavior in 2D, where
the onset of the cluster growth is very smooth and depends 
on dissipation $\diss$ only implicitly.

When cluster growth has reached a dynamic equilibrium in the
{\em saturation regime}, the system is dominated by one large cluster 
which contains a macroscopic fraction of the system.
Note that this regime still has an interesting dynamics and
smaller clusters interact with the large cluster. Neither the
small clusters nor the large one are static and the latter loses or
eats up particles or smaller clusters.
Kinetic energy and collision frequency still fluctuate, 
but are governed by the equations $E(\tau) \sim \tau^{-2}$ and 
$f_c(\tau) \sim \tau^{-1}$.  This means the evolution in time is 
similar to the homogeneous cooling state.

A very interesting observation is the similarity of clustering in 
3D and percolation with respect to the critical exponents.
Even for the dynamics of cluster growth we have found the same exponents 
as for the occupation probability in percolation theory.
We have provided a tentative explanation which is based on the linear
relation of the occupation probability $p$ and time $\tau$ around 
the onset $\tg$ of cluster growth in 3D.
In 2D we did not find a power law behavior because the growth of the large
cluster happens when the linear relation between $p$ 
and $\tau$ has disappeared\,---\,at least for the given set 
of parameters.

However, there still remain some open questions:
Why does $\tg$ depend on $r$ in 3D, but not in 2D?
Why does the growth of the large cluster proceed smoothly in 2D and rather 
sharply in 3D?  And why are all these differences present while 
the energy decay is proportional to $\tau^{-1}$ in both cases? 
Detailed studies of these questions and also of the 
cluster size probability distribution are in progress as 
well as a more systematic study of the cluster definition in 3D.

\begin{acknowledgments}
This research was supported by the DFG, the SFB 382, the SFB 404, 
and LU 450/9-1.
We thank Hans Herrmann and Sean McNamara for helpful discussions.
\end{acknowledgments}

\bibliography{bib/ed,bib/percol,bib/cluster,bib/tmp,bib/granular,bib/own}

\begin{thebibliography}{23}
\expandafter\ifx\csname natexlab\endcsname\relax\def\natexlab#1{#1}\fi
\expandafter\ifx\csname bibnamefont\endcsname\relax
  \def\bibnamefont#1{#1}\fi
\expandafter\ifx\csname bibfnamefont\endcsname\relax
  \def\bibfnamefont#1{#1}\fi
\expandafter\ifx\csname citenamefont\endcsname\relax
  \def\citenamefont#1{#1}\fi
\expandafter\ifx\csname url\endcsname\relax
  \def\url#1{\texttt{#1}}\fi
\expandafter\ifx\csname urlprefix\endcsname\relax\def\urlprefix{URL }\fi
\providecommand{\bibinfo}[2]{#2}
\providecommand{\eprint}[2][]{\url{#2}}

\bibitem[{\citenamefont{Herrmann et~al.}(1998)\citenamefont{Herrmann, Hovi, and
  Luding}}]{herrmann98}
\bibinfo{editor}{\bibfnamefont{H.~J.} \bibnamefont{Herrmann}},
  \bibinfo{editor}{\bibfnamefont{J.-P.} \bibnamefont{Hovi}}, \bibnamefont{and}
  \bibinfo{editor}{\bibfnamefont{S.}~\bibnamefont{Luding}}, eds.,
  \emph{\bibinfo{title}{Physics of dry granular media - NATO ASI Series E 350}}
  (\bibinfo{publisher}{Kluwer Academic Publishers},
  \bibinfo{address}{Dordrecht}, \bibinfo{year}{1998}).

\bibitem[{\citenamefont{P\"oschel and Luding}(2001)}]{poschel01}
\bibinfo{editor}{\bibfnamefont{T.}~\bibnamefont{P\"oschel}} \bibnamefont{and}
  \bibinfo{editor}{\bibfnamefont{S.}~\bibnamefont{Luding}}, eds.,
  \emph{\bibinfo{title}{Granular Gases}} (\bibinfo{publisher}{Springer},
  \bibinfo{address}{Berlin}, \bibinfo{year}{2001}), \bibinfo{note}{lecture
  Notes in Physics 564}.

\bibitem[{\citenamefont{Vermeer et~al.}(2001)\citenamefont{Vermeer, Diebels,
  Ehlers, Herrmann, Luding, and Ramm}}]{vermeer01}
\bibinfo{editor}{\bibfnamefont{P.~A.} \bibnamefont{Vermeer}},
  \bibinfo{editor}{\bibfnamefont{S.}~\bibnamefont{Diebels}},
  \bibinfo{editor}{\bibfnamefont{W.}~\bibnamefont{Ehlers}},
  \bibinfo{editor}{\bibfnamefont{H.~J.} \bibnamefont{Herrmann}},
  \bibinfo{editor}{\bibfnamefont{S.}~\bibnamefont{Luding}}, \bibnamefont{and}
  \bibinfo{editor}{\bibfnamefont{E.}~\bibnamefont{Ramm}}, eds.,
  \emph{\bibinfo{title}{Continuous and Discontinuous Modelling of Cohesive
  Frictional Materials}} (\bibinfo{publisher}{Springer},
  \bibinfo{address}{Berlin}, \bibinfo{year}{2001}), \bibinfo{note}{lecture
  Notes in Physics 568}.

\bibitem[{\citenamefont{Kishino}(2001)}]{kishino01}
\bibinfo{editor}{\bibfnamefont{Y.}~\bibnamefont{Kishino}}, ed.,
  \emph{\bibinfo{title}{Powders \& Grains 2001}} (\bibinfo{publisher}{Balkema},
  \bibinfo{address}{Rotterdam}, \bibinfo{year}{2001}).

\bibitem[{\citenamefont{Cafiero et~al.}(2000)\citenamefont{Cafiero, Luding, and
  Herrmann}}]{cafiero00}
\bibinfo{author}{\bibfnamefont{R.}~\bibnamefont{Cafiero}},
  \bibinfo{author}{\bibfnamefont{S.}~\bibnamefont{Luding}}, \bibnamefont{and}
  \bibinfo{author}{\bibfnamefont{H.~J.} \bibnamefont{Herrmann}},
  \bibinfo{journal}{Phys. Rev. Lett.} \textbf{\bibinfo{volume}{84}},
  \bibinfo{pages}{6014} (\bibinfo{year}{2000}).

\bibitem[{\citenamefont{Goldhirsch and Zanetti}(1993)}]{goldhirsch93}
\bibinfo{author}{\bibfnamefont{I.}~\bibnamefont{Goldhirsch}} \bibnamefont{and}
  \bibinfo{author}{\bibfnamefont{G.}~\bibnamefont{Zanetti}},
  \bibinfo{journal}{Phys. Rev. Lett.} \textbf{\bibinfo{volume}{70}},
  \bibinfo{pages}{1619} (\bibinfo{year}{1993}).

\bibitem[{\citenamefont{Goldhirsch et~al.}(1993)\citenamefont{Goldhirsch, Tan,
  and Zanetti}}]{goldhirsch93b}
\bibinfo{author}{\bibfnamefont{I.}~\bibnamefont{Goldhirsch}},
  \bibinfo{author}{\bibfnamefont{M.-L.} \bibnamefont{Tan}}, \bibnamefont{and}
  \bibinfo{author}{\bibfnamefont{G.}~\bibnamefont{Zanetti}},
  \bibinfo{journal}{Journal of Scientific Computing}
  \textbf{\bibinfo{volume}{8}}, \bibinfo{pages}{1} (\bibinfo{year}{1993}).

\bibitem[{\citenamefont{McNamara and Young}(1996)}]{mcnamara96}
\bibinfo{author}{\bibfnamefont{S.}~\bibnamefont{McNamara}} \bibnamefont{and}
  \bibinfo{author}{\bibfnamefont{W.~R.} \bibnamefont{Young}},
  \bibinfo{journal}{Phys. Rev. E} \textbf{\bibinfo{volume}{53}},
  \bibinfo{pages}{5089} (\bibinfo{year}{1996}).

\bibitem[{\citenamefont{Luding and Herrmann}(1999)}]{luding99}
\bibinfo{author}{\bibfnamefont{S.}~\bibnamefont{Luding}} \bibnamefont{and}
  \bibinfo{author}{\bibfnamefont{H.~J.} \bibnamefont{Herrmann}},
  \bibinfo{journal}{Chaos} \textbf{\bibinfo{volume}{9}}, \bibinfo{pages}{673}
  (\bibinfo{year}{1999}).

\bibitem[{\citenamefont{Nie et~al.}(2002)\citenamefont{Nie, Ben-Naim, and
  Chen}}]{nie02}
\bibinfo{author}{\bibfnamefont{X.~B.} \bibnamefont{Nie}},
  \bibinfo{author}{\bibfnamefont{E.}~\bibnamefont{Ben-Naim}}, \bibnamefont{and}
  \bibinfo{author}{\bibfnamefont{S.~Y.} \bibnamefont{Chen}},
  \bibinfo{journal}{Phys. Rev. Lett.} \textbf{\bibinfo{volume}{89}}
  (\bibinfo{year}{2002}).

\bibitem[{\citenamefont{Das and Puri}(2003{\natexlab{a}})}]{das03a}
\bibinfo{author}{\bibfnamefont{S.~K.} \bibnamefont{Das}} \bibnamefont{and}
  \bibinfo{author}{\bibfnamefont{S.}~\bibnamefont{Puri}},
  \bibinfo{journal}{Physica A} \textbf{\bibinfo{volume}{318}},
  \bibinfo{pages}{55} (\bibinfo{year}{2003}{\natexlab{a}}).

\bibitem[{\citenamefont{Das and Puri}(2003{\natexlab{b}})}]{das03b}
\bibinfo{author}{\bibfnamefont{S.~K.} \bibnamefont{Das}} \bibnamefont{and}
  \bibinfo{author}{\bibfnamefont{S.}~\bibnamefont{Puri}},
  \bibinfo{journal}{Europhys. Letters} \textbf{\bibinfo{volume}{61}},
  \bibinfo{pages}{749} (\bibinfo{year}{2003}{\natexlab{b}}).

\bibitem[{\citenamefont{Luding}(1998)}]{luding98cref}
\bibinfo{author}{\bibfnamefont{S.}~\bibnamefont{Luding}}, in
  \emph{\bibinfo{booktitle}{Physics of dry granular media - NATO ASI Series
  E350}}, edited by \bibinfo{editor}{\bibfnamefont{H.~J.}
  \bibnamefont{Herrmann}}, \bibinfo{editor}{\bibfnamefont{J.-P.}
  \bibnamefont{Hovi}}, \bibnamefont{and}
  \bibinfo{editor}{\bibfnamefont{S.}~\bibnamefont{Luding}}
  (\bibinfo{publisher}{Kluwer Academic Publishers},
  \bibinfo{address}{Dordrecht}, \bibinfo{year}{1998}), p. \bibinfo{pages}{285},
  \bibinfo{note}{also see references therein}.

\bibitem[{\citenamefont{Lubachevsky}(1991)}]{lubachevsky91}
\bibinfo{author}{\bibfnamefont{B.~D.} \bibnamefont{Lubachevsky}},
  \bibinfo{journal}{J.~Comput.~Phys.} \textbf{\bibinfo{volume}{94}},
  \bibinfo{pages}{255} (\bibinfo{year}{1991}).

\bibitem[{\citenamefont{Miller and Luding}(2004)}]{sm:par}
\bibinfo{author}{\bibfnamefont{S.}~\bibnamefont{Miller}} \bibnamefont{and}
  \bibinfo{author}{\bibfnamefont{S.}~\bibnamefont{Luding}},
  \bibinfo{journal}{J.~Comp. Phys.} \textbf{\bibinfo{volume}{193}},
  \bibinfo{pages}{306} (\bibinfo{year}{2004}), \bibinfo{note}{physics/0302002}.

\bibitem[{\citenamefont{Allen and Tildesley}(1987)}]{allen87}
\bibinfo{author}{\bibfnamefont{M.~P.} \bibnamefont{Allen}} \bibnamefont{and}
  \bibinfo{author}{\bibfnamefont{D.~J.} \bibnamefont{Tildesley}},
  \emph{\bibinfo{title}{Computer Simulation of Liquids}}
  (\bibinfo{publisher}{Oxford University Press}, \bibinfo{address}{Oxford},
  \bibinfo{year}{1987}).

\bibitem[{\citenamefont{Lubachevsky}(1992)}]{lubachevsky92}
\bibinfo{author}{\bibfnamefont{B.~D.} \bibnamefont{Lubachevsky}},
  \bibinfo{journal}{Int.~J.\ Comput.\ Simul.} \textbf{\bibinfo{volume}{2}},
  \bibinfo{pages}{372} (\bibinfo{year}{1992}).

\bibitem[{\citenamefont{Luding and McNamara}(1998)}]{luding98f}
\bibinfo{author}{\bibfnamefont{S.}~\bibnamefont{Luding}} \bibnamefont{and}
  \bibinfo{author}{\bibfnamefont{S.}~\bibnamefont{McNamara}},
  \bibinfo{journal}{Granular Matter} \textbf{\bibinfo{volume}{1}},
  \bibinfo{pages}{113} (\bibinfo{year}{1998}),
  \bibinfo{note}{cond-mat/9810009}.

\bibitem[{\citenamefont{Luding et~al.}(1998)\citenamefont{Luding, Huthmann,
  McNamara, and Zippelius}}]{luding98d}
\bibinfo{author}{\bibfnamefont{S.}~\bibnamefont{Luding}},
  \bibinfo{author}{\bibfnamefont{M.}~\bibnamefont{Huthmann}},
  \bibinfo{author}{\bibfnamefont{S.}~\bibnamefont{McNamara}}, \bibnamefont{and}
  \bibinfo{author}{\bibfnamefont{A.}~\bibnamefont{Zippelius}},
  \bibinfo{journal}{Phys. Rev. E} \textbf{\bibinfo{volume}{58}},
  \bibinfo{pages}{3416} (\bibinfo{year}{1998}).

\bibitem[{\citenamefont{Haff}(1983)}]{haff83}
\bibinfo{author}{\bibfnamefont{P.~K.} \bibnamefont{Haff}}, \bibinfo{journal}{J.
  Fluid Mech.} \textbf{\bibinfo{volume}{134}}, \bibinfo{pages}{401}
  (\bibinfo{year}{1983}).

\bibitem[{\citenamefont{Sela and Goldhirsch}(1995)}]{sela95}
\bibinfo{author}{\bibfnamefont{N.}~\bibnamefont{Sela}} \bibnamefont{and}
  \bibinfo{author}{\bibfnamefont{I.}~\bibnamefont{Goldhirsch}},
  \bibinfo{journal}{Phys. Fluids} \textbf{\bibinfo{volume}{7}},
  \bibinfo{pages}{507} (\bibinfo{year}{1995}).

\bibitem[{\citenamefont{Stauffer and Aharony}(1994)}]{stauffer94}
\bibinfo{author}{\bibfnamefont{D.}~\bibnamefont{Stauffer}} \bibnamefont{and}
  \bibinfo{author}{\bibfnamefont{A.}~\bibnamefont{Aharony}},
  \emph{\bibinfo{title}{Introduction to Percolation Theory}}
  (\bibinfo{publisher}{Taylor \& Francis}, \bibinfo{address}{London},
  \bibinfo{year}{1994}), \bibinfo{edition}{2nd} ed.

\bibitem[{\citenamefont{Tobochnik}(1999)}]{tobochnik99}
\bibinfo{author}{\bibfnamefont{J.}~\bibnamefont{Tobochnik}},
  \bibinfo{journal}{Phys. Rev. E} \textbf{\bibinfo{volume}{60}},
  \bibinfo{pages}{7137} (\bibinfo{year}{1999}).

\end{thebibliography}

\end{document}